\documentclass[traditabstract]{aa} 
\usepackage{graphicx,longtable}
\usepackage{txfonts,multirow}

\def\ltsima{$\; \buildrel < \over \sim \;$}
\def\lsim{\lower.5ex\hbox{\ltsima}}
\def\gtsima{$\; \buildrel > \over \sim \;$}
\def\gsim{\lower.5ex\hbox{\gtsima}}
\def\mdot {\dot M}
\newcommand{\be}{\begin{equation}}
\newcommand{\en}{\end{equation}}

\def\rsole {~R_{\odot}}
\def\msole {~M_{\odot}}

\def\lae{\mathrel{<\kern-1.0em\lower0.9ex\hbox{$\sim$}}}
\def\gae{\mathrel{>\kern-1.0em\lower0.9ex\hbox{$\sim$}}}

\begin{document}

\title{A universal relation for the propeller mechanisms in magnetic rotating stars at different scales }

\author{
Sergio Campana\inst{1}, Luigi Stella\inst{2}, Sandro Mereghetti\inst{3}, Domitilla de Martino\inst{4}}
\institute
{INAF - Osservatorio Astronomico di Brera, Via E. Bianchi 46, I-23807, Merate (LC), Italia \\ \email{sergio.campana@brera.inaf.it}
\and
INAF - Osservatorio Astronomico di Roma, Via Frascati 33, I-00078, Monteporzio Catone (Roma), Italia
\and
INAF - Istituto di Astrofisica Spaziale e Fisica Cosmica - Milano, Via E. Bassini 15, I-20133, Milano, Italia
\and
INAF - Osservatorio Astronomico di Capodimonte, Salita Moiariello 16, I-80131 Napoli, Italia
}

\date{}

\abstract{
Accretion of matter onto a magnetic, rotating object can be strongly affected by the interaction with its magnetic field.
This occurs in a variety of astrophysical settings involving young stellar objects, white dwarfs, and neutron stars.
As matter is endowed with angular momentum, its inflow toward the star is often mediated by an accretion disc. 
The pressure of matter and that originating from the stellar magnetic field
balance at the magnetospheric radius: at smaller distances  the 
motion of matter is dominated by the magnetic field, and funnelling towards the magnetic poles ensues.
However, if the star, and thus its magnetosphere, is fast spinning, most of the inflowing matter will be halted 
at the magnetospheric radius by centrifugal forces, resulting in a characteristic reduction of the accretion luminosity.
The onset of this mechanism, called the propeller, has been widely adopted to interpret 
a distinctive knee in the decaying phase of the 
light curve of several transiently accreting X--ray pulsar systems.
By comparing the observed luminosity at the knee for different classes of objects with the value predicted by 
accretion theory on the basis of the independently measured
magnetic field, spin-period, mass, and radius of the star, we disclose here a general relation for the onset of the propeller 
which spans about eight orders of magnitude in spin period and ten in magnetic moment. 
The parameter-dependence and normalisation constant that we determine are in agreement with basic accretion theory. 
}


\authorrunning{S. Campana et al.} 

\maketitle

\section{Introduction}

The innermost regions of accretion flows towards a magnetic star are governed by the magnetic field which channels the
inflowing material onto the magnetic polar caps. The motion of the inflowing matter starts being dominated by the magnetic field at the 
magnetospheric radius, where the star's magnetic field pressure balances the ram pressure of the infalling matter.
In spherical symmetry, simple theory predicts that this will  occur at the   Alfv\'en radius,
$$
r_{\rm A}=\left( {{\mu^4}\over{2\,G\,M\,\mdot^2}} \right)^{1/7}, \eqno{(1)}
$$
where $\mu$ is the magnetic dipole moment ($\mu=B\,R^3$), $\mdot$ is the mass accretion rate, $G$ is the gravitational 
constant, and $M$, $R$, and $B$ are the star mass, radius, and dipolar magnetic field, respectively 
(Pringle \& Rees 1972; Frank, King \& Raine 2002).
In interacting binaries, the accreting matter is endowed with a large angular momentum, which leads to  
the formation of an accretion disc where matter spirals in by transferring angular momentum outwards, 
while gravitational  energy is converted into internal energy and/or radiation.
The interaction of an accretion disc with a magnetosphere has been investigated  
extensively  through theoretical studies 
(Ghosh \& Lamb 1979; Wang 1987; Shu et al. 1994; 
Lovelace, Romanova \& Bisnovatyi-Kogan 1995; Li \& Wickramasinghe 1997; 
D'Angelo \& Spruit 2012; Parfrey, Spitkovsky \& Belobodorov 2016) 
and through magnetohydrodynamical simulations 
(Romanova et al. 2003; Zanni \& Ferreira 2013).

\begin{figure*}
\begin{center}
\includegraphics[width=.9\textwidth]{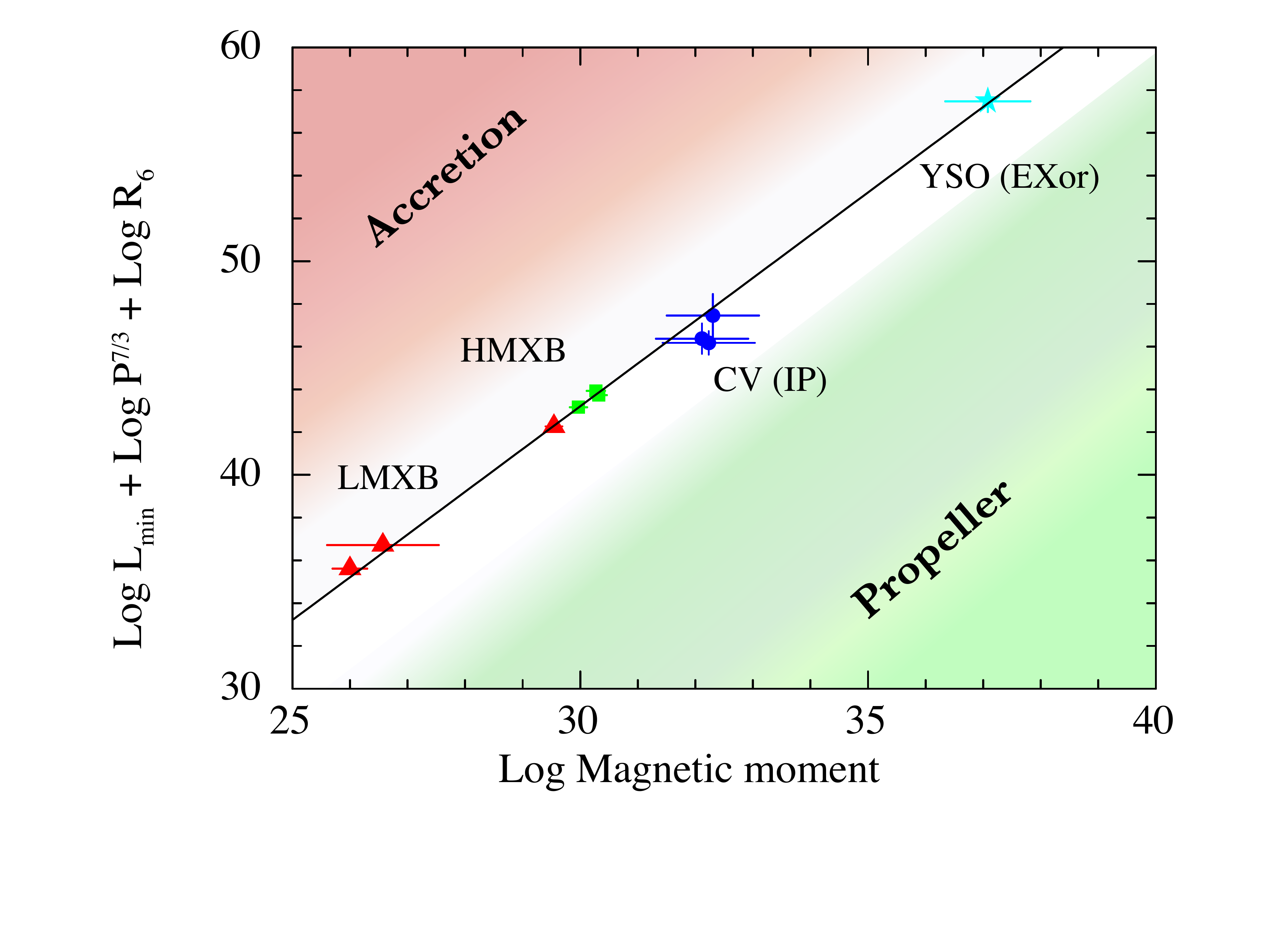}
\caption{Correlation among the star magnetic dipole $\mu$ (in Gauss cm$^3$ units) estimated with different methods 
(unrelated to the accretion theory) and a combination of stellar parameters $P^{7/3}\,R$ ($P$ in seconds and $R$ in units 
of $10^6$ cm) at the observationally determined luminosity {at which the outburst light curves shows a knee or a drop}
($L_{\rm min}$ in erg s$^{-1}$). 
We indicated LMXBs with red triangles, HMXBs with green squares, 
white dwarfs in intermediate polars with blue dots, and YSOs with cyan stars. The continuous line represents the best fit power law relation
and separates the region where unimpeded matter accretion onto the star surface takes place 
from the region where it is halted  (partially or completely)  by the propeller mechanism.}
\label{Fig1}
\end{center}
\end{figure*}

In some models (Ghosh \& Lamb 1979; Shu et al. 1994; Wang 1987)   
the boundary between the disc and the magnetosphere is expressed through 
a slight modification of Eq. (1): the dependence on the parameters remains the same, 
but the magnetospheric radius is $r_{\rm m}=\xi\,r_{\rm A}$, with $\xi\sim 0.1-1$ a normalisation 
factor that depends on the assumptions of different theories 
(Ghosh \& Lamb 1979; Spruit \& Taam 1993; Wang 1987; Kluzniak \& Rappaport 2007).
Models which consider different types of disc magnetic-field threading  and 
outflow-launching mechanisms also lead to a similar dependence on the parameters 
(Shu et al. 1994; Lovelace, Romanova \& Bisnovatyi-Kogan 1995; Li \& Wickramasinghe 1997).
Other studies predict  instead  a different dependence of the magnetospheric radius  on the stellar
parameters. For example,  three-dimensional magnetohydrodynamical simulations 
(Romanova \& Owocki 2016), 
predict a flatter dependence than 
$\mu^{4/7}$, while the dead-disc model yields a steeper dependence 
(D'Angelo \& Spruit 2012).

Disc matter orbiting at Keplerian speeds, where truncation by magnetic stresses occurs, may rotate 
faster or slower than the magnetosphere, depending on whether the corotation radius
$$
r_{\rm cor}=\left( {{G\,M\,P^2}\over{4\,\pi^2}}\right)^{1/3} \eqno{(2)}
$$
is larger or smaller than the magnetospheric radius, respectively ($P$ is the rotational period of the star).
In the former case 
(i.e. $r_{\rm m}<r_{\rm cor}$) gravity dominates and matter can enter the magnetosphere, 
get attached to the field lines, and 
proceed toward the star surface in a near free fall.
In the  case where $r_{\rm m}>r_{\rm cor}$, 
the centrifugal drag by the magnetosphere exceeds gravity, and most 
matter is propelled away or halted at the boundary, rather than being accreted. 
This centrifugal inhibition induces a marked reduction of the 
accretion-generated luminosity (by a factor of up to $\sim r_{\rm m}/R$ for radiatively efficient
accretion); it was originally proposed as an explanation
for the small number of luminous 
Galactic X--ray binary systems (mainly accreting neutron stars, 
Illarionov \& Sunyaev 1975). 
Stella et al. (1986) discussed the propeller mechanism 
in the context of accreting X--ray pulsar 
systems undergoing large variations of the mass inflow rate and showed that
the sharp steepening in the decaying X--ray luminosity of transient
systems (notably the case of Be-star system V 0332+53) could be interpreted  
as being due to the onset of centrifugal inhibition of accretion. 
In subsequent studies, the same mechanism was applied 
to interpret the sharp decay, usually taking place over a few 
days, in the light curves of both 
high mass and low mass X--ray binary systems hosting a 
rotating magnetic neutron star (e.g. Aql X-1, Campana et al. 1998; 4U 0115+63 Campana et al. 2001). 
 Values of the neutron star magnetic field in these systems derived from
the application of the propeller mechanism 
were found to be broadly consistent with those inferred through
cyclotron resonant features.
The propeller mechanism may also be expected to operate in 
all other classes of magnetic, rotating objects undergoing accretion 
and displaying outbursts that involve substantial changes in 
the mass inflow rate. 
In fact the condition for the onset of the propeller ($r_{\rm m} \sim r_{\rm cor}$) is close to the spin equilibrium 
condition. Spin equilibrium can be attained by a rotating magnetic star secularly as a result of the spin-up and 
spin-down torques operating on the magnetic star (Ghosh \& Lamb 1979; Priedhorsky 1986; 
Yi, Wheeler \& Vishniac 1997; Nelson et al. 1997; Campbell 2011; Parfrey, Spitkovsky \& Beloborodov 2016).
Different regimes can be probed 
through the luminosity variations of up to several orders of magnitude
induced by the propeller,
depending on the mass inflow rate and the star's spin period and magnetic field strength 
(Campana et al. 1998).

\begin{table*}
\caption{Summary of  the sources used in Fig. 1.}
{\footnotesize
\begin{center}
\begin{tabular}{ccccccc}
\hline
Source                      &Log Dipole moment   &Log Spin period& Log Radius &Mass          &  Log Lim. luminosity  & Distance\\
                                   & (G cm$^3$)                 & (s)                       &(cm)                &($M_\odot$)& (erg s$^{-1}$)             & (kpc)\\
\hline
{\bf LMXBs}              &                                      &                            &                         &                   &                                       &\\
SAX J1808.4--3658&$26.0^{+0.3}_{-0.3}$ & $-2.60$             & 6.00               & 1.4            &$35.8^{+0.1}_{-0.2}$ & 3.5\\
IGR J18245--2452  &$26.6^{+1.0}_{-1.0}$ & $-2.41$             & 6.00               & 1.4            &$36.3^{+0.2}_{-0.3}$  & 5.5\\
GRO J1744-28        &$29.5^{+0.1}_{-0.1}$ & $-0.33$             & 6.00               &  1.4           &$36.9^{+0.1}_{-0.1}$ & 8.0\\
\hline
{\bf HMXBs}              &                                      &                            &                        &                   &                                        &\\
4U 0115+63             &$30.0^{+0.2}_{-0.2}$ & $0.56$              & 6.00               & 1.4            &$35.8^{+0.2}_{-0.3}$ & 7.0\\
V 0332+53                &$30.3^{+0.2}_{-0.2}$  & $0.64$              & 6.00               & 1.4            &$36.3^{+0.1}_{-0.1}$ & 7.0\\
SMC X-2                   &$30.3^{+0.1}_{-0.1}$ & $0.37$              & 6.00                & 1.4            &$36.8^{+0.1}_{-0.1}$ & 62.0\\
\hline
{\bf CVs}                  &                                      &                            &                          &                   &                                          &\\
WZ Sge                    &$32.1^{+0.8}_{-0.8}$ & 1.46                   & 8.77                &  0.85          & $34.2^{+0.7}_{-0.7}$ & 0.04\\
V455 And                 &$32.3^{+0.8}_{-0.8}$ & 1.83                   & 8.81                &  0.60         & $33.2^{+0.5}_{-0.5}$ & 0.09\\             
HT Cam                    &$32.2^{+0.8}_{-0.8}$ & 2.71                   & 8.83               &   0.55         & $32.5^{+1.0}_{-1.0}$ & 0.10\\
\hline
{\bf YSOs}                 &                                        &                            &                       &                   &                                            &\\
V1647 Ori                 &$37.1^{+0.7}_{-0.7}$   & 4.94                   &   11.45          & 0.80         & $34.6^{+0.5}_{-0.5}$ & 0.40\\  
\hline
\end{tabular}
\end{center}
\noindent Magnetic moments for WDs are slightly different due to the different radii of the compact objects.
Errors are at the $1\,\sigma$ confidence level.\\}
\end{table*}

\section{Propeller transition in different classes of sources}

We consider here different classes of accretion-powered transient sources 
that possess a disc, and search for signs of the
transition from the accretion to the propeller regime in their light curves:
these are neutron stars in low mass and high mass X--ray binaries
(LMXBs and HMXBs), white dwarfs of the intermediate polar (IP) class  displaying dwarf nova outbursts, 
and young stellar objects in FUor or EXor-like objects. 
In our sample we include only accreting stars with observationally determined 
(or constrained) values of the magnetic field, in addition to
measurements of the spin period, distance, mass and radius.
These quantities are required to determine the dependence 
of the limiting luminosity before the onset of the propeller 
(see below and Section 6).
Therefore, we exclude objects in which the magnetic field is inferred 
only on the basis of phenomena related to the interaction 
between magnetosphere and accretion flow (e.g. accretion torques).
The resulting suitable objects are  a small sample, but they provide the key to  
testing the relationship
between the magnetospheric radius and the star's parameters. 
In the following sections we describe in detail the values of the physical 
parameters that we adopted for each  selected source. 

We took a purely observational approach and initially
inspected the outburst light curves of transient sources with known parameters, 
searching for a sharp knee or drop.\footnote{While we cannot rule out 
that   knees or drops are caused in some cases by other mechanisms  
(e.g. disc instabilities) than the action of the propeller, most sources in our sample 
have physical parameters that make them close to 
the propeller condition. This issue is further discussed in Section 6.}
Then, if a knee or a drop was found in the outburst light curve,
we worked out the corresponding luminosity and retained the source in the sample. 

If these drops are interpreted in terms of the propeller mechanism, simple modelling predicts that the accretion luminosity $L_{\rm acc}=G\,M\,\mdot/R$ will decrease to 
$L_{\rm mag}=G\,M\,\mdot/r_{\rm m}$ after the drop, as the inflow of matter is halted at the magnetospheric radius ($r_{\rm m}>R$)  
by the centrifugal barrier (Corbet 1996; Campana et al. 1998). 
The limiting luminosity before the onset of the propeller is obtained by equating the magnetospheric radius to the corotation radius, resulting in 
$$
L_{\rm lim}\sim 1.97\times 10^{38}\,\xi^{7/2}\,\mu_{30}^2\,P^{-7/3}\,M^{-2/3}\,R_6^{-1} . \eqno{(3)}
$$
Here $\mu$ is in $10^{30}$ G cm$^3$, $P$ in seconds, $M$ in solar masses, and $R$ in $10^6$ cm units.
Simulations, as well as observations of a few transitional accreting millisecond pulsars 
at the onset of the propeller, show that the mass inflow is not halted 
completely by the propeller;  some matter still finds its way towards the star surface
(Ustyugova et al. 2006; Romanova \& Owocki 2016). 

Observationally two different behaviours are found: outbursts with a marked decay (or knee)  in the light curve and  
outbursts with a smoother light curve decay. For the former we calculate the limiting luminosity $L_{\rm lim}$ 
as the bolometric luminosity at the onset of the steep decay. We evaluate bolometric corrections
on a case-by-case basis  (see  below).
For the sources displaying a smooth decay, we assume that unimpeded accretion of matter 
onto the surface takes place even in quiescence. 
In our framework a smooth decay would imply that the source does not enter the propeller regime, and thus an
indirect constraint on the neutron star parameters.

\section{Observations and results: neutron stars}

In this section we discuss in some detail the accreting magnetic neutron stars in high and low mass X--ray binaries which are in our sample. 
Special attention is given to the determination of the magnetic fields. 
In particular, the magnetic field strength for the two accreting millisecond pulsars in LMXBs 
is derived from the radio-pulsar spin-down measured during quiescence 
(Hartman et al. 2009; Papitto et al. 2013) and for the four HMXBs from the 
cyclotron resonant scattering features in their X--ray spectra (Tsygankov et al. 2016).

\subsection{SAX J1808.4--3658}

SAX J1808.4--3658, the first accreting millisecond pulsar discovered 
(Wijnands \& van der Klis 1998), has a spin period of 2.5 ms.
The magnetic moment was estimated by interpreting the spin-down during quiescence as being 
due to rotating magnetic dipole radiation; this gave 
$\mu=9.4\times10^{25}$ G cm$^{3}$ 
(Hartman et al. 2009). 
Other methods have been suggested in the literature, but all of them involve the magnetospheric radius and so are not relevant to the present
study. We note, however, that the value of $\mu$ obtained through these estimates is broadly consistent with 
the value derived from rotating magnetic dipole radiation.
Errors and values have been symmetrised in log scale for fitting purposes, and we adopted $\log(\mu/{\rm G\ cm^{3}})=26.0^{+0.3}_{-0.3}$.

Gilfanov et al. (1998) after the discovery outburst already noted a fast dimming of the source around a luminosity of $\sim 5\times 10^{35}$ erg s$^{-1}$.
Swift monitoring observations of the 2005 outburst led Campana, Stella \& Kennea  (2008) 
to estimate a limiting 0.3--10 keV luminosity of $(2.5\pm1.0)\times
10^{35}$ erg s$^{-1}$ by taking the lowest value during outburst and the luminosity of the rebrightenings.
Using the broad-band Suzaku spectrum during outburst 
(Cackett et al. 2009), 
we estimated a bolometric correction of 1.7, giving a 0.1--100 keV
limiting luminosity of $\log(L_{\rm lim}/{\rm  erg\ s^{-1}})=35.8^{+0.1}_{-0.2}$.

\subsection{IGR J18245--2452}

IGR J18245--2452 is a transitional pulsar, the only one so far found to alternate between the accretion powered regime and
 the millisecond radio pulsar regime (Papitto et al. 2013).
IGR J18245--2452 lies in the M28 globular cluster at 5.5 kpc. The spin period is 3.9 ms (Papitto et al. 2013).
Also in this case the magnetic field can be derived from the spin-down in the radio pulsar regime. However, the period derivative is 
known with a large uncertainty and it is likely affected by the source acceleration in M28.
Ferrigno et al. (2014) 
set a limit on the magnetic field to $(0.7-35)\times 10^8$ G.
This results in a magnetic dipole moment of $\log(\mu/{\rm G\ cm^3})=26.6^{+1.0}_{-1.0}$. 
The Swift monitoring of the 2013 outburst showed a steepening in the X--ray light curve around a 0.5--10 keV
 luminosity of $\sim 10^{36}$ erg s$^{-1}$ (Linares et al. 2014). 
An XMM-Newton observation caught IGR J18245--2452  
continuously switching between a low and a 
high intensity state, with flux variations of a factor of $\sim 100$ on timescales of a few seconds 
(Ferrigno et al. 2014). 
This behaviour was interpreted in terms of centrifugal inhibition of accretion. 
Pulsations have also been detected  at lower luminosities, though with pulse fraction decreasing with luminosity. 
On the contrary at higher luminosities the pulse fraction remained
nearly constant. By adopting a bolometric correction of 2.5 based on spectral fits, we derived
a limiting 0.1--100 keV luminosity of $\log(L_{\rm lim}/{\rm  erg\ s^{-1}})=36.3^{+0.2}_{-0.3}$.

\begin{table*}
\caption{Swift/XRT observation log for GRO J1744--28.}
\footnotesize{
\begin{center}
\begin{tabular}{cccrr}
\hline
Observation        &Duration  &Date             &Counts$^*$& Pulsed Fraction\\
(00030898XXX)&(ks)            &                      & (Burst)   &           ($\%$)$^\#$ \\
\hline
045                       &0.9            &2014-04-21&20737 (Y)& $9.4\pm1.6$ \\
046                       &1.0            &2014-04-23&25017 (N)& $8.5\pm1.5$ \\
047                       &0.9            &2014-04-26&13853 (Y)& $8.7\pm1.9$ \\
048                       &1.0            &2014-04-27&16165 (N)& $9.6\pm1.7$ \\
049                       &1.1            &2014-04-29&17039 (Y)& $8.9\pm1.7$ \\              
050                       &1.0            &2014-05-01&11297 (N)& $9.0\pm2.1$ \\
051                       &2.0            &2014-05-05&14543 (N)& $10.8\pm1.9$ \\
052                       &0.4            &2014-05-07&  2079 (N)& $6.1\pm4.8$ \\
053                       &1.8            &2014-05-09&  7268 (N)& $12.5\pm2.6$ \\
054                       &1.4            &2014-05-11&  5200 (N)& $4.1\pm3.1$ \\
056                       &1.3            &2014-05-19&  1842 (N)& $2.7\pm5.3$ \\
057                       &1.7            &2014-05-23&  1675 (N)& $2.2\pm5.8$ \\
058$^+$              &1.1            &2014-05-25&    693 (N)& $11.3\pm8.5$ \\
059                       &1.9            &2014-05-27&  1223 (N)& $2.6\pm6.3$ \\
060$^+$              &1.7            &2014-05-29&    913 (N)& $19.0\pm7.0$ \\
\hline
\end{tabular}
\end{center}
{\leftline{$^*$ Counts extracted from a 20 pixel radius region. Data were all taken in Window Timing mode.}}
{\leftline{$^\#$ Pulsed fractions were computed folding the data (20 bin) at the $P=0.467045$ s period and fitted with a single  }}
{\leftline{sinusoidal function  plus constant.  Errors are at the $90\%$ confidence level.}}
{\leftline{$^+$ In these folded light curves there is just one point  which drives the fit, resulting in a large pulsed fraction.}}}
\end{table*}

\begin{figure}
\begin{center}
\includegraphics[width=0.5\textwidth]{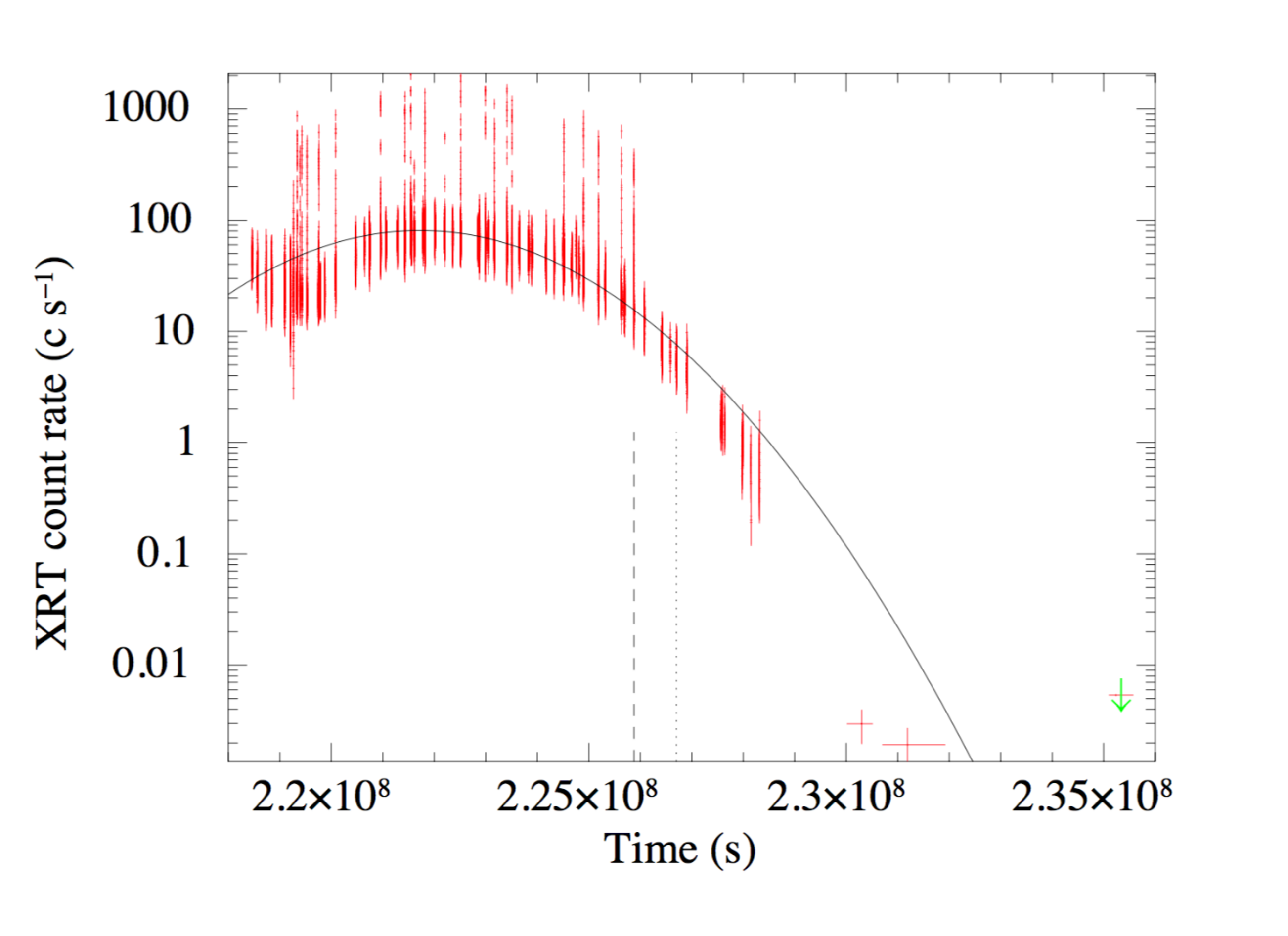}
\caption{Swift/XRT light curve of the 2014 outburst of GRO 1744--28. 
Variations in the XRT count rate along each observation are apparent, and the strong flares are due to Type II burst activity (thought to be due
to  accretion instability at the magnetosphere). The continuous line shows a Gaussian
fit to the first part of the outburst (excluding the flaring activity). This fit does not apply to the latest stages of the outburst, which display a steeper decay.
The dashed vertical line marks the time at which Type II bursting activity disappeared. The dotted vertical line marks the time at which pulsations
were not detected.} 
\label{Fig2}
\end{center}
\end{figure}

\subsection{GRO J1744--28}

GRO J1744--28 is a peculiar transient bursting pulsar discovered close to the Galactic Centre 
(Kouveliotou et al. 1996). 
It has a pulse period of 0.467 s (Finger et al. 1996).
The magnetic field has been directly estimated 
from an absorption cyclotron feature at $E_{\rm cyc} \sim 4.7\pm0.1$ keV in the X--ray spectrum with 
a hint of higher order harmonics 
(D'A\`i et al. 2015; see also Doroshenko et al. 2015 who found a slightly lower line energy at $\sim 4.5$ keV).
Its centroid energy is related to the magnetic field through $E_{\rm cyc} =11.6\times (B/10^{12} {\rm G})/(1+z)$ keV, where $z$ is the
gravitational redshift at the neutron star surface. 
Here and throughout this work we assume values of $1.4\msole$ and 10 km for the mass and radius of the neutron
star, respectively.
For these values the gravitational redshift is $z=0.21$,
and a magnetic field of $B=(4.9\pm0.1)\times 10^{11}$ G is derived, corresponding to a magnetic dipole moment of
$\log(\mu/{\rm G\ cm^3})=29.5^{+0.1}_{-0.1}$.
GRO J1744--28 is likely associated with the Galactic Centre population at a distance of 8 kpc.
A drop in the X--ray flux of a factor of $\sim 10$ and the disappearance of the pulsed signal in the RXTE data were interpreted by Cui 
(1997) as the onset of the propeller regime. 
The 2--60 keV absorbed limiting flux has been estimated as  $\sim 2.3 \times 10^{-9}$ erg cm$^{-2}$ s$^{-1}$. 
Given the power law spectrum provided by Cui (1997, 1998), 
we derive a bolometric correction of 1.5 for the 0.1--100 keV unabsorbed flux.
This results in a limiting luminosity of $\log(L_{\rm lim}/{\rm  erg\ s^{-1}})\sim 37.4$.

This result is, however, subject to possible systematic uncertainties due to the high crowding in the Galactic centre region observed by  the 
non-imaging RXTE proportional counter array.
To improve on this, we take advantage of the deep monitoring of GRO 1744--28 carried out with the Swift satellite (Gehrels et al. 2004) 
during the 2014 outburst. 
The Swift/XRT (Burrows et al. 2005) 
light curve can be found in Fig. A1 (which was derived through the  Leicester University automatic pipeline, 
Evans et al. 2009\footnote{http://www.swift.ac.uk/user\_objects/}). 
During the main phase of the outburst the light curve was affected by a large number of bright bursts that characterise the 
`bursting pulsar' GRO 1728--44.
These bursts ceased  abruptly when the count rate decreased below $\sim 20$ c s$^{-1}$. 
After a few days  the periodic pulsations, which had a pulsed fraction of $10\pm2\%$ ($90\%$ confidence level), 
also dropped to $<8\%$ (see Table 1). Fitting the pulsed fraction before and after obsid. 00030898054  (see Table 2) with the same value, 
results in an unacceptable fit ($3\,\sigma$).
Fitting the outburst light curve with a (phenomenological) Gaussian model, from the peak to the time when the pulsations 
become undetectable, shows that at later times 
there is a hint of a steeper turn off of the observed rate than model extrapolation (a Lorentzian or an exponential
function provides poorer fits).
Together with the disappearance of bursting activity and of X--ray pulsations, this can be taken as an indirect
evidence for the propeller onset, even though differences from 
the abrupt turn off observed in other systems are apparent.

We estimate the limiting luminosity from the Swift spectrum of the first observation when no pulsations are detected 
(obsid. 00030898054, see Table 1). 
The source spectrum was extracted with a 20 pixel radius circle, whereas the background spectrum was extracted at the 
edge of the Window Timing (WT) strip. Energy-resolved data were rebinned to 1 count per channel and fitted 
with an absorbed power law model by using C-statistics.
We found $N_H=(9.8\pm1.1)\times 10^{22}$ cm$^{-2}$ and a hard power law with  
$\Gamma=1.2\pm0.2$, consistent with previous estimates 
(Younes et al. 2015).
The 0.3--10 keV unabsorbed flux is $9.5\times10^{-10}$ erg cm$^{-2}$ s$^{-1}$, 
which, extrapolated to the 0.1--100 keV range, results in $\log(L_{\rm lim}/{\rm  erg\ s^{-1}})=36.9^{+0.1}_{-0.1}$, 
a factor of 3 lower than that estimated by Cui (1997).

\subsection{4U 0115+63}

4U 0115+63 is a well-known hard X--ray transient binary, hosting a 3.61 s spinning neutron star orbiting an O9e companion (V635 Cas; 
Rappaport et al. 1978).
Cyclotron features were observed in the X--ray spectrum during outbursts (White, Swank \& Holt 1983);
four cyclotron harmonics were revealed during a BeppoSAX observation, resulting in a precise determination of the fundamental cyclotron 
line energy of $E_{\rm cyc} =12.74\pm0.08$ keV (Santangelo et al. 1999). 
Using the same redshift as before, this converts to a magnetic moment of $\log(\mu/{\rm G\ cm^3})=30.0^{+0.1}_{-0.2}$. 
The distance was estimated to be $\sim 7$ kpc  (Negueruela et al. 2001). 
4U 0115+63 showed a very large variation  during a long BeppoSAX observation (a factor of $\sim 250$ in 15 hr; Campana et al. 2001), which was interpreted in terms of the action of the centrifugal barrier.
A knee was present in the rising phase (see Fig. 2 intervals 2 and 3 in 
Campana et al. 2001) from which we estimate a limiting luminosity in the 0.1--100 keV 
of $\log(L_{\rm lim}/{\rm  erg\ s^{-1}})=35.8^{+0.2}_{-0.3}$. 
Our estimate is consistent with the one by Tsygankov et al. (2016) 
evaluated through a drop in the Swift/XRT light curve of the 2015 outburst, even though the latest stages of the outburst were not well sampled.

\subsection{V 0332+53}

V 0332+53 is a hard X--ray transient similar to 4U 0115+63. The companion is an O8-9e-type main sequence star at a distance 
of $\sim 7$ kpc (Negueruela et al. 1999). 
The neutron star spin period is 4.4 s (Doroshenko, Tsygankov \& Santangelo 2016). 
Cyclotron line features were discovered by Makishima et al. (1990). 
Detailed studies carried out with RXTE and INTEGRAL revealed up to two higher harmonics and a fundamental 
line energy between 25--27 keV depending on the modelling, (Kreykenbohm et al. 2005; Pottschmidt et al. 2005). 
This leads to a redshift corrected magnetic dipole moment $\log(\mu/{\rm G\ cm^3})=30.3^{+0.2}_{-0.2}$.
Stella et al. (1986) 
were the first to find signs of centrifugal inhibition of accretion in the outburst decay of V 0332+53.
A detailed analysis by Tsygankov et al. 
(2016b) 
of the 2015 outburst decay with Swift/XRT showed a rapid turn off at a limiting 
luminosity of $\log(L_{\rm lim}/{\rm  erg\ s^{-1}})=36.3^{+0.1}_{-0.1}$ (0.1--100 keV).

\subsection{SMC X-2}

The Small Magellanic Cloud ($d=62$ kpc) hosts a large number of Be-star X--ray binaries. One of them, 
SMC X-2, has a short spin period of 2.37 s (Corbet et al. 2001). 
NuSTAR and Swift simultaneous observations showed the presence of a cyclotron line 
feature at $28\pm1$ keV (Jaisawal \& Naik 2016). 
This leads to a redshift corrected magnetic dipole moment $\log(\mu/{\rm G\ cm^3})=30.3^{+0.1}_{-0.1}$.
A dense Swift/XRT monitoring of the recent 2016 outburst (Lutovinov et al. 2017) reveals a steepening 
in the outburst light curve at a luminosity of  $\sim 4\times 10^{36}$ erg s$^{-1}$.
Assuming a correction factor of 1.5 to account for our lower energy bound (from 0.5 to 0.1 keV), 
we obtain  $\log(L_{\rm lim}/{\rm  erg\ s^{-1}})=36.8^{+0.1}_{-0.1}$.

\subsection{Other sources}

We note that the abrupt X--ray flux drop of more than a decade in minutes to hours that was 
observed in several HMXB pulsators with spin periods in the $\sim 100$ s 
range was ascribed to a sudden decrease in the density of the 
companion's wind from which they accrete 
(Vela X-1, Kreykenbohm et al. 2009; GX 301-2, Gogus et al. 2011; 4U 1907+09, Doroshenko et al. 2012); 
therefore, we did not include these sources in our sample.
Other HMXBs for which the magnetic field was measured from cyclotron line features possess  a spin period that is too long to 
propel the inflowing matter away during any stage of their outbursts.

The two transitional millisecond X--ray pulsars PSR J1023+0038 (Archibald et al. 2009) and XSS J12270--4859 (de Martino et al. 2013; 
Bogdanov et al. 2015; Roy et al. 2015), which have known spin periods, magnetic fields, and distances, 
have not yet displayed an X--ray outburst and therefore are not included in the sample.

\section{Observations and results: white dwarfs}

Accreting magnetic white dwarfs are usually classified as  intermediate polars and polars
depending on the strength of the magnetic field and, in turn, whether or not  a 
disc forms. 
The white dwarf magnetic field in polars is strong enough to prevent the formation of an accretion disc and they do not display outbursts. 
On the other hand, a small number of intermediate polars display dwarf  nova-like outbursts. 
These sources are generally short orbital period systems 
(i.e. $P_{\rm orb} < 2$ hr) except for GK Per ($P_{\rm orb}\sim 48$ hr) and very few others.
We selected 68 intermediate polars from Mukai's web page\footnote{http://asd.gsfc.nasa.gov/Koji.Mukai/iphome/iphome.html},  
including all sources flagged as probable, confirmed, and ironclad. 
We further down-selected these by keeping only white dwarfs with
known spin period, mass, and distance estimates.
Radii were computed from $R/R_\odot=0.008\ (M/M_\odot)^{-1/3}$ (Nauenberg 1972). 

Magnetic fields are difficult to measure  due to the lack of detectable optical/near-IR polarisation
in the majority of IPs. A few of them were found to be polarised with loosely estimated B-fields $\sim (4-30)\times 10^6$ G 
(see Ferrario et al. 2015 and 
references therein). We therefore assume a conservative upper limit of $4\times 10^6$ G for the systems under study
simply because none of the sources under the present analysis has shown hints of optical/near-IR polarisation yet.

A lower limit cannot be easily worked out. A non-negligible magnetic field is needed since the emission is pulsed and thus matter 
is channeled onto the poles.
Based on this we considered two intermediate polars showing pulsed emission in X--rays at high luminosities: YY (DO) Dra
(Szkody et al. 2002) and GK Per (Yuasa et al. 2016).
From the best bremsstrahlung model fitting of the X--ray spectrum of these sources
we derived the density of the accreting material (assuming a conservative 
column height of the order of the stellar radius). 
Then, equating the thermal pressure to the magnetic pressure we derived a magnetic field in excess of $\sim 10^5$ G in both cases. 
The same value is obtained by calculating the limiting magnetic field
for which column-like accretion would still occur over the entire white dwarf surface
(Frank, King \& Raine 2002).
Therefore we conservatively assume  $B>10^5$ G as a lower limit for the magnetic field in intermediate polars.

\subsection{WZ Sge}

WZ Sge is a dwarf nova showing regular outbursts with a recurrence period of $\sim 30$ yr. It shows super-outbursts (7 mag amplitude) 
with characteristic reflaring at the end of the primary event.
WZ Sge has an orbital period of 81.6 min and lies 
at a distance of 43 pc based on parallax measurements. The mass of the white dwarf is $0.85\msole$ 
(Steeghs et al. 2007). 
Several oscillations around 28 s have been discovered in the optical. Optical observations in quiescence revealed a 27.87 s signal that was 
phase stable from night to night 
(Patterson 1980). 
In addition a signal at the same period was detected in the 2--6 keV range from ASCA data, but not in previous 
X--ray observations (Patterson et al. 1998),  
which likely represents the white dwarf spin period. 
WZ Sge is a good candidate IP  (Lasota et al. 1999).
A distinctive knee in the 2001 outburst light curve was observed around $V\sim 11$ 
(Kato et al. 2009). 
Ultraviolet observations allowed an order of magnitude estimate of the mass accretion rate before and after the time of the transition.
During the plateau phase, before the occurrence of the knee, the
mass accretion rate was $\mdot\sim (1-3)\times 10^{-9}\msole$ yr $^{-1}$ 
(Long et al. 2003), 
whereas it was $\mdot\sim 5\times 10^{-10}\msole$ yr$^{-1}$ at the peak of the first rebrightening,  
close in magnitude to the occurrence of the knee  
(Godon et al. 2004).
Taking the mean value of these two estimates and increasing the error by a factor of two, we end up with 
$\log(L_{\rm lim}/{\rm  erg\ s^{-1}})=34.2^{+0.7}_{-0.7}$.

It is important to note that WZ Sge is the prototype of a subclass of dwarf novae characterised by optical light curves which 
display large outbursts, followed by a knee in the decay phase and subsequent rebrightenings 
(Kato 2015). 
WZ Sge is one of the few members of this class for which the spin period has been inferred. 

Despite other interpretations of dimmings and rebrightenings during dwarf nova outburst decays
(e.g. Meyer \& Meyer-Hofmeister 2015; Hirose \& Osaki 1990), which can still occur but in the end will produce a decrease in the mass inflow rate and in turn the 
onset of the propeller, here the inference of our work is that all WZ Sge-type binaries should host a fast spinning white dwarf.
Based on the knee luminosity and magnetic field in the above-described range, we expect spin periods
in the 10--500 s range.

\subsection{V455 And}

V455 And (also known as HS 2331+3905) is a unique cataclysmic variable that was discovered in the Hamburg Quasar Survey.
Partial eclipses in the optical light curve recur at the orbital period of 81.1 min. A stable period at 67.62 s 
was found in quiescent optical data, which is interpreted as the spin period of the 
white dwarf (Bloemen et al. 2012). 
Modelling the quiescent far-UV emission of the white dwarf leads to a distance of 90 pc 
(Araujo-Betancor et al. 2005).  
No firm estimate of the mass has been obtained; we assumed $0.6\msole$, the same value used to fit the far-UV  spectrum.

In 2007 September, V455 And increased its brightness by 8 mag, similar to the outburst amplitudes observed in WZ Sge stars.
A step-like (2 mag) decrease was observed $\sim 20$ d after the beginning of the outburst when V455 And had $V=12.7$.
Multifilter observations allowed Matsui et al. (2009) 
to estimate a temperature of $(10-12)\times 10^3$ K at the transition, and we derive 
a bolometric correction of a factor of 10  with respect to the observed $V$ magnitude.

\begin{table*}
\label{table:upper}
{\footnotesize
\begin{center}
\caption{Summary table for white dwarfs in confirmed IPs with smooth outburst declines, providing upper limits.}
\begin{tabular}{ccccccc}
\hline
Source                      &Log Dipole moment   &Log Spin period& Log Radius & Mass        & Log Limiting luminosity& Distance\\
                                   & (G cm$^3$)                 & (s)                       & (cm)              &($M_\odot$)& (erg s$^{-1}$)                &(kpc)\\
\hline
{\it AE Aqr}                 &$32.8^{+1.0}_{-1.0}$   &  1.52                  & 8.78              &0.8            &  $>31.0^{+0.3}_{-0.3}$& 0.1\\
\hline
{\it GK Per }              &$32.1^{+0.8}_{-0.8}$ & 2.55                     & 8.77              &   0.87         & $<33.7^{+0.3}_{-0.3}$ & 0.47\\
{\it YY (DO) Dra }    &$32.1^{+0.8}_{-0.8}$ & 2.72                     &8.77                & 0.84           & $<32.1^{+0.4}_{-0.4}$ & 0.16\\
{\it V1223 Sgr}         &$32.1^{+0.8}_{-0.8}$ & 2.87                    & 8.77               & 0.87           & $<34.1^{+0.3}_{-0.3}$ & 0.53\\ 
{\it TV Col}               &$32.0^{+0.8}_{-0.8}$ & 3.28                     & 8.75               &    1.00         & $<33.9^{+0.3}_{-0.3}$ & 0.37\\
{\it EX Hya  }            &$32.1^{+0.8}_{-0.8}$ & 3.60                    & 8.78                &     0.79       & $<32.4^{+0.4}_{-0.4}$ & 0.07\\
\hline
\end{tabular}
\end{center}
Errors are at the $1\,\sigma$ confidence level.}
\end{table*}

\subsection{HT Cam}

HT Cam is a cataclysmic variable with a short orbital period (81 min). 
Its 515.1 s spin period was found from optical photometric and X--ray data  (Kemp et al. 2002). 
Following de Martino et al. (2005), 
we assume a white dwarf mass of $0.55\msole$ and a distance of 100 pc.

HT Cam shows rare and brief outbursts from $V\sim 17.8$ to $V\sim 12-13$. One of these outbursts
was monitored in detail in 2001 by Ishioka et al. (2002). 
After a short plateau, HT Cam showed a dramatic decline of more than 4 mag d$^{-1}$. 
The decline rate changed around $V\sim 14$, when  pulsations also became less pronounced 
(Ishioka et al. 2002).
It is difficult 
to derive the luminosity at the turning point since the transition to quiescence was extremely rapid (a few days) and no
multiwavelength data are available. We therefore consider the bolometric quiescent luminosity as derived from the X--ray band  
of $9\times 10^{30}$ erg s$^{-1}$  (de Martino et al. 2005) 
and correct it for the luminosity change in the optical by 3.8 mag. 
We adopt in this case a total error of an order of magnitude (Table 1).

\subsection{AE Aqr}

AE Aqr is an intermediate polar which hosts one of the fastest  known white dwarfs (33 s spin
period) in an IP system (Patterson 1979).

It shows a peculiar phenomenology, with strong broad-band variability from the radio to X--ray and possibly TeV energies. 
Doppler tomography based on optical spectra showed an
irregular accretion pattern that has been interpreted 
as arising from the action of the propeller mechanism (Wynn, King \& Horne 1997).
The distance is $\sim 100$ pc  (based on Hipparcos data) and the mass 
has been estimated to be $0.8\msole$ (Casares et al. 1996).
Estimates for the magnetic field are uncertain. Values in the $(0.3-30)\times 10^6$ G
have been obtained, as summarised by Li et al. (2016).
The X--ray luminosity of $10^{31}$ erg s$^{-1}$, provides a reasonable estimate
of the bolometric luminosity (Oruru \& Meintjes 2012; see also the recent Swift/XRT+NuSTAR estimate 
Kitaguchi et al. 2014, Table 3).

\subsection{Other sources}

There are a  number of other IPs which do not display 
hints of a knee in their outburst light curve decay (see Tables 1 and 3).
Consequently, we assume that their quiescent emission 
is powered by accretion onto the white dwarf surface and  adopt the corresponding
luminosity as an upper limit to $L_{\rm lim}$. Figure 3 shows these upper limits 
plotted together with the points in Fig. 1; all of them are consistent with the propeller line, while others are  close to it.

Owing to insufficient data (convincing distance, mass estimates, and more importantly a sound bolometric correction for the knee luminosity), 
CC Scl was not included in our sample even though it showed a drop in flux that could be due to the action of the propeller.

\begin{table}[t]
\label{table:wd_para}
\footnotesize{
\begin{center}
\caption{White dwarf parameters.}
\begin{tabular}{cccc}
\hline
Source                     &  Limiting mag      &Black-body $T$         & Refs.\\
                                  &                               &                                     & \\
\hline
WZ Sge$^\#$          & $V\sim 11$          & from UV                      & 1\\  
V455 And$^\&$       &  $V=12.7$            & from opt/near-IR                     & 2\\   
HT Cam$^*$           &$V\sim 14.5$        & from X--rays               & 3\\ 
\hline
\hline
Source                          &  Peak mag      &  Quiescent               & Refs.\\
                                       &                          &      mag                      & \\
\hline
{\it GK Per}$^*$           & $V\sim10.2$   & $V\sim13.5$             &4, 5\\
{\it YY (DO) Dra}$^\#$&$V\sim 10$    & $V\sim 15.5$           & 6, 7 \\
{\it V1223 Sgr}$^*$     &$V\sim 12$     & $V\sim13.5$            & 4, 8\\
{\it TV Col}$^*$            & $V\sim11$     & $V\sim14$                & 9, 6 \\
{\it EX Hya$^*$}           & $V\sim 10$    & $V\sim13.5$            & 10, 6 \\
\hline
\end{tabular}
\end{center}
$\#$ Modelling of the UV emission with an accretion disc.\\
$^\&$ Bolometric correction $10\pm1$.\\
$^*$ Magnetic cataclysmic variable, X--ray flux dominates the emission. \\
Refs. 1: Godon et al. 2004; 
2: Matsui et al. 2009;%
3: de Martino et al. 2005;
4: Landi et al. 2009;
5: Barlow et al. 2006;
6: Yuasa et al. 2010;
7: Hoard et al. 2005;
8: Hayashi \& Ishida 2014;
9: Vrtilek et al. 1996;
10: Luna et al. 2015.}
\end{table}

\begin{figure*}
\begin{center}
\includegraphics[width=0.8\textwidth,angle=-90]{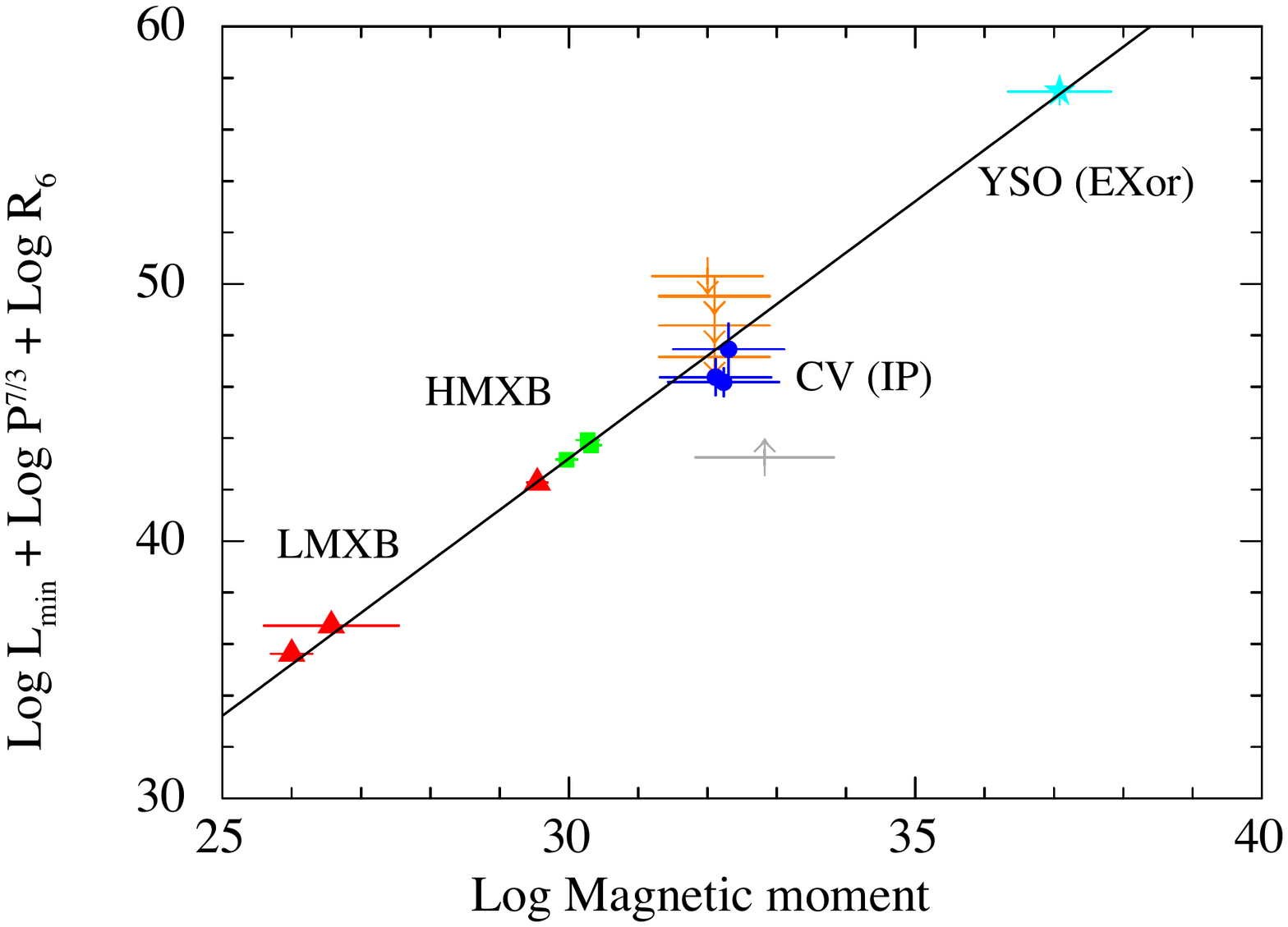}
\vskip -1.8truecm
\caption{As in Fig. 1. We have included here a few upper limits from WDs in IPs close to the propeller line (orange upper limits),
and one source (AE Aqr) considered to persistently lie in the propeller regime (grey lower limit). The upper and lower limits are displayed to show
the consistence with the relation, but they were not used in the fit.}
\label{Fig3}
\end{center}
\end{figure*}

\section{Observations and results: young stellar objects}

Young stellar objects (YSOs) are often surrounded by an accretion disc formed 
by matter infalling from the parent cloud.
A small number of these low mass pre-main sequence stars show large outbursts 
or eruptions of several magnitudes, signifying 
episodes of enhanced mass accretion rate.
Based on their properties, they are classified as FUors (named after the prototype of this class FU Ori),
which display violent and probably recurrent outbursts about once per century, and EXors 
(EX Lup ancestor), which show shorter duration (weeks to months) and 
less pronounced irregular outbursts several times in a decade
(Hartmann, Herczeg \& Calvet 2016; Audard et al. 2014).
FUors and EXors are relatively rare; for only one YSO, V1647 Ori, we could find all 
values that are needed for our study. 
EX Lup itself might have shown a steep drop in flux at the end of the 2008 outburst, 
but the steepening did not have a knee; instead  
its light curve morphology resembles the dip-like structure seen in nova outburst from DQ Her 
and likely caused by the formation of dust.
FU Ori initiated a major outburst from 15--16 mag to about 10 mag  over several months
in 1936 with a very slow decay and no sign of a turn off. 
The mass accretion rate increased by 2--3 orders of magnitude
(Beck \& Aspin 2012); 
therefore, we did not include these YSOs in our sample.

\subsection{V1647 Ori}

V1647 Ori is an eruptive YSO in Orion. It showed several outbursts that do not share the timescales and 
amplitudes of EXors.
Evidence for accretion were clearly found in spectroscopic data as testified by the Br $\gamma$ emission line 
(Brittain et al. 2010). 
It went into outburst in 2004-2006 and then again in 2008. Together with an increase in the optical and near-IR magnitudes,
V1647 Ori displayed a correlated bright outburst in the X--ray band 
(Teets et al. 2011). 
Based on X--ray outburst data Hamaguchi et al. (2012) detected a 1~d periodicity, identified  as the protostar rotation at near 
break-up speed. The emission is interpreted as originating at the magnetic footprints of an accretion flow governed by
magnetic reconnection.
The shape of the 1~d X--ray periodicity was found to be consistent with the modulation
expected from two antipodal hot spots, suggesting that accretion took place through a
mainly dipolar field geometry (see Appendix A).
Therefore, rotation takes place at near break-up speed in V1647 Ori.
In order to confine the 50 MK plasma,
as derived from X--ray spectral fits, a magnetic field of at least 100 G must be invoked 
(Hamaguchi et al. 2012).
As an upper limit on $B$ we take the mean field measured in isolated YSOs 
of $3\times 10^3$ G (Johns-Krull 2007).
Based on pre-outburst bolometric luminosity and stellar temperature, Aspin et al. (2008) 
estimated a mass of $\sim 0.8\msole$ and a radius of $\sim 5\rsole$ for V1647 Ori at a distance of 400 pc.

V1647 Ori emits preferentially in the near-IR. The $H$-band light curve showed a clear steepening at a luminosity of $\sim 8\times 10^{32}$
erg s$^{-1}$ (Teets et al. 2011). 
However, the luminosity at the knee is not easy to estimate. In addition, V1647 Ori is heavily absorbed.
We started from the estimate of the mass accretion rate during quiescence of $\sim 5\times 10^{-7}\msole$ yr$^{-1}$ based on 
the flux of the Br$\gamma$ emission line (Acosta-Pulido et al. 2007). 
By using the same method for the outburst maximum, Acosta-Pulido et al. (2007) 
derived a peak mass accretion rate of $\sim 5\times 10^{-6}\msole$ yr$^{-1}$. 
A factor of $\sim 10$ variation in the mass inflow rate is in agreement with the $H$-band light curve, which 
showed an enhancement of a factor of $\sim 10$ in the near-IR flux 
(Hamaguchi et al. 2012). 
Given that the knee occurred close to the middle of the light curve in log scale, we took the logarithmic mean between 
the two mass accretion rates (quiescent and peak outburst) and converted it to a bolometric luminosity
by using $G\,M\mdot/R$. A logarithmic error of $\pm0.3$ was assumed (see Table 1).

\section{A universal relation for the limiting luminosity}

The values of  $L_{\rm lim}$ obtained from our sample 
span about 5 decades; they are listed in Table 1 (see also Fig. 1).
We carried out a parametric search 
in order to determine the dependence of $L_{\rm lim}$ on the star's parameters and test the 
theory of disc-magnetosphere interaction.
We adopt a general relation of the form 
$$
L_{\rm lim}\sim k\,\mu^{\alpha}\,P^{-\beta}\,R^{-\gamma} , \eqno{(4)}
$$
similar to Eq. (3) but with free exponents to be derived from the fit to the observational data. 
We performed a grid search on $\alpha$, $\beta$, and $\gamma$ over the interval from  0 to 4 for 
each parameter, a larger range than envisaged in all models.
We also included a fixed mass dependence 
to the $-2/3$ power (as in Eq. 3) because  the relatively small range
of stellar masses (less than a factor of 3, spanning $0.55-1.4\msole$) does not allow us to 
test  this dependence independently.
We proceeded as follows: we selected a value on the grid of the exponents, establish the best fitting relation, i.e. determine $k$,
and took note of the $\chi^2$ of the fit. 
Based on this grid search, we found the best fit (reduced $\chi^2=0.97$ for 9 degrees of freedom)
and determined the $90\%$ confidence 
intervals for three interesting parameters.
We obtained $\alpha=1.9^{+0.6}_{-0.3}$,  $\beta=2.3^{+0.8}_{-0.4}$, and  
$\gamma=1.1^{+0.3}_{-0.3}$ (we note that there is an almost linear correlation between $\alpha$ and $\beta$, 
which is accounted for in the evaluation of the errors). 
The values of all parameters are consistent  
with the dependence predicted by simple disc-magnetosphere interaction theory (Eq. 3) 
over a very wide range of objects and physical occurrences.

We can also use our sample with exponents fixed at $\alpha=2$,  $\beta=7/3$, and  $\gamma=1$ 
in order to evaluate the 
$\xi$ parameter from the best  fit value of $k$ (see Eqs. (3) and (4)). We obtain $\xi=0.49\pm0.05$ ($1\,\sigma$), a value 
consistent with that predicted in some disc-magnetosphere models 
\footnote{In some works the magnetic dipole moment is defined as $\mu=B_p\,R^3/2$, 
where $B_p$ is the field at the magnetic poles (as opposed to the magnetic equator). In this case $\xi=0.66\pm0.07$.} 
(Ghosh \& Lamb 1979; Wang 1987).
We note that the precision attained in the determination of the parameters  stems from
the very wide range of magnetic dipole moments ($> 10$ decades), spin periods 
($> 7$ decades), and star radii ($>5$ decades) in our sample.
We thus conclude that our study provides the most extensive observational 
confirmation of basic theory for the accretion disc-magnetosphere interaction.

In deriving our conclusions we implicitly assumed that the same accretion 
disc-magnetosphere interaction holds for different astrophysical objects,
an approach adopted  in other studies 
(Romanova \& Owocki 2016 and references therein).
Even if we cannot prove this assumption, 
our findings are consistent with 
different classes of sources in our sample behaving similarly 
and in a consistent manner with the propeller mechanism (see Section 3).

An entirely different scenario based on a DIM 
has been widely adopted to model the flux drops observed 
in the outburst decay of black hole transients (and dwarf 
novae)\footnote{Black hole systems do not possess a magnetosphere 
and thus the propeller mechanism cannot operate.}. 
The disc mass inflow rate decreases on an exponential decay timescales of a few weeks, 
$\tau \sim 40\,R_{\rm h,11}^{5/4}$ d ($R_{\rm h,11}$ the maximum hot-state disc radius 
in units of $10^{11}$ cm; King \& Ritter 1988).
In these systems the matter proceeds unimpeded all the way to the compact object.
After the transition to the cold state the 
outburst light curves should show a characteristic `brink' (i.e. a knee), followed 
by a linear decay (Powell, Haswell \& Falanga 2007).

In the best sampled light curves the decay after the knee is not linear but exponential 
(e.g. Aql X-1 Campana et al. 2014 or 4U 0115+63 Campana et al. 2001).
Campana, Coti Zelati and D'Avanzo (2013) studied 20 outbursts from Aql X-1,
showing that the brink when detectable always occurred at a luminosity larger than the 
knee luminosity. Unfortunately, the less sampled sources cannot clearly reveal these differences.

The decrease in luminosity across the knee in the HMXBs 
(e.g. V 0332+53 and 4U 0115+63, which have spins of several seconds
and $r_{\rm cor} \gg R$ ) is $\sim 10^2$, whereas it is much smaller in 
LMXBs, CVs, and YSOs (where $r_{\rm cor}$ is only a few times $R$).
This has no obvious interpretation in the DIM, where the luminosity
swing is determined only by the accretion rate in the two states
of the disc; on the contrary, it is naturally interpreted by the 
propeller model, where the luminosity swing is expected to scale 
approximately\footnote{When $r_{\rm cor}$ is close to $R$ 
it is easier for some matter to leak through the centrifugal barrier, 
further reducing the luminosity gap and possibly making the transition 
smoother.} as $r_{\rm cor} /R$ (Corbet 1996; Campana et al 1998).
 
Moreover, if the sharp decay in the light curve of one or more of our sources were 
due to the DIM (or to some other mechanism that does not block the matter flow), 
the corresponding points would be in the red region above the correlation line in Fig.~1. 
We do not observe any obvious outliers confirming that our interpretation 
applies to most  if not all sources in our sample.

The fact that the relation expressed in Eq.~(4) (derived from the correlation in Fig.~1)
corresponds approximately to the condition $r_{\rm m} \sim r_{\rm cor}$ 
also indicates  that the present-day values of the star's
spin and magnetic field have been largely determined by its 
accretion history (and consequent spin-up/spin-down torques
and/or $B-$field decay) over evolutionary timescales.  
Therefore, the correlation itself cannot result from 
angular momentum or magnetic flux conservation 
in the star's progenitor. 

Given our parametrisation of Eq.~(4) we can rewrite the magnetospheric radius (Eq.~1) as
$$
r_{\rm m}=k\,\left(\mu^{\alpha}\,L^{-\gamma}\,R^{-\gamma} \right)^{{2}\over{3\,\beta}}, \eqno{(5)}
$$
which is independent of the central mass object.
For standard accretion theory values $\alpha=2$, $\beta=7/3$, and $\gamma=1$ we derive 
$$
r_{\rm m}=(3.3\pm0.5)\times 10^8\, \left(\mu_{30}^{4}\,L_{36}^{-2}\,R_6^{-2}\,M\right)^{{1}\over{7}} \ {\rm cm}, \eqno{(6)}
$$
with $L_{36}$ the accretion luminosity in units of $10^{36}$  erg s$^{-1}$.
Eq. (3) instead becomes 
$$
L_{\rm lim}\sim 1.65^{+0.94}_{-0.60}\times 10^{37}\,\mu_{30}^2\,P^{-7/3}\,M^{-2/3}\,R_6^{-1}  \ {\rm erg \ s^{-1}}. \eqno{(7)}
$$

\section{Conclusions}

We determined the critical luminosity at which the transition from accretion to propeller regime occurs in a number of 
transient neutron stars, white dwarf, and young stellar objects, whose spin period, radius, and magnetic field are 
observationally measured (or constrained).
The transition luminosity was estimated on a pure observational basis as the luminosity 
at which the outburst light curve shows a distinctive knee. This allowed us to probe the 
dependence of this 
critical luminosity on the star's main parameters, independently of the accretion theory. 
We found good agreement of the best fit indices and those predicted by accretion theory
if the knee arises from the transition to the propeller regime. 

The observationally tuned Eq. (3), together with the value of $\xi \simeq 0.5$ that we 
determined, adds confidence and precision to previous studies exploiting the knee associated to the onset of the centrifugal 
barrier to infer (or constrain) the magnetic field of accreting neutron stars of known spin period. 
For instance the magnetic field of the prototypical LMXB transient, Aql X-1, can be estimated as $\mu= (5.7\pm0.3)
\times 10^{26}$ G cm$^3$ (at 4.5 kpc) based on the luminosity of the knee observed in the outburst light curves in 2010 and 2014 
(Campana et al. 2014). The same is true for a number of fast spinning neutron stars in HMXBs in the Magellanic Clouds 
(Christodoulou et al. 2016).
Moreover Eq. (3) extends the applicability of the method and its predictive power 
to other classes of objects, notably magnetic white dwarfs in CVs and YSOs.
There is an emerging class of short orbital period  dwarf novae showing
steep decays in their optical light curves, often followed by rebrightenings.
These are known as WZ Sge dwarf novae (Kato 2015).
The prototypical source, WZ Sge, is also a candidate IP   occasionally displaying fast pulsations at 28 s. 
For all sources of this class we predict similar characteristics, i.e. a magnetic field in the $10^5-10^6$ G range and a fast 
spin period in the 10--500 s range. 
In the study of FUor and EXor, drops in the near-IR light curves,  coupled with estimates of the spin period based on X--ray 
and optical/near-IR polarimetry, will provide a new tool for evaluating their magnetic fields.

\begin{acknowledgements}
We thank F. Coti Zelati, P. D'Avanzo, T. Giannini, D. Lorenzetti, and B. Nisini for useful conversations.
DDM and LS acknowledge support from ASI-INAF contract I/037/12/0.
Swift is a NASA mission with participation of the Italian Space Agency and the UK Space Agency. 
This research has made use of software and tools provided by the High Energy Astrophysics Science 
Archive Research Center (HEASARC), which is a service of the Astrophysics Science Division at NASA/GSFC 
and the High Energy Astrophysics Division of the Smithsonian Astrophysical Observatory.
This work has made use of data supplied by the UK Swift Science Data Centre at the University of Leicester.
\end{acknowledgements}

\appendix
\section{Multipole magnetic fields}

Virtually all models of the disc-magnetospheric interaction, 
assume that the star's magnetic field is dipolar. The possibility that only higher multipole components 
are present  can be ruled out since the dependence on the parameters in Eq. (1) would be 
inconsistent with the values that we determined. 
For instance, for a pure quadrupolar field it would have a radial dependence $B_{\rm q}\sim \mu_{\rm q}/r^5$, where
$\mu_{\rm q}$ is the magnetic quadrupole moment.
The expected scaling of the magnetospheric radius in this case would be $\mu_{\rm q}^{4/15}\,\mdot^{-2/15}$ 
(Vietri \& Stella 1998). This is excluded by our analysis (see Section 6).

An implicit requirement of our study is that the dipolar component dominates not only 
at the disc-magnetosphere boundary, but also at the radius where the magnetic field 
value is estimated (even if higher multipole components were present, the dipole 
would still dominate at large radii by virtue of its weaker radial dependence).
Among neutron stars only magnetars and a few accreting pulsars in ultraluminous X--ray 
sources are believed to feature conspicuous higher multipole components at the star surface 
(Tiengo et al. 2013; Israel et al. 2017). 
However, no evidence for such components has ever been found in the HMXB neutron stars in our sample,
for which the magnetic field is estimated through cyclotron resonant features.
A quadrupolar magnetic field component has been suggested for Her X-1 to model the 
complex pulse profile (Panchenko \& Postnov 1994); however, its contribution to the magnetic field 
should be only $\sim 30\%$ (Klochkov et al. 2008) at the neutron star surface (as measured from the cyclotron line) 
and negligible at the magnetospheric radius (which is $\sim10^8$~cm farther away). 
Only a weak dipolar magnetic field is expected to survive in the fast spinning neutron stars of LMXBs.
The magnetic field estimate that we adopt is based on 
the spin-down torque in the quiescent (no-accretion) phase and thus it exclusively involves 
 the dipole component of the field.
Higher magnetic multipole components are observed in young white dwarfs, but in older systems like short
period binaries ($\gsim 1$ Gyr) a substantial decay of multipolar components is expected (Beuermann et al. 2007).
In YSOs surface magnetic fields which comprise 
contributions from higher magnetic multipoles have been found  
(Johns-Krull 2007). 
However, in the case of V1657 Ori the shape of the rotationally modulated X--ray
light curve argues in favour of magnetic accretion through a dipole-dominated field.

We thus conclude that our  assumption that the magnetic dipole dominates over higher multipole components appears 
to be well justified for the objects in our sample.

\end{document}